# The rise and fall of Mott insulating gaps in YNiO$_3$ paramagnets as a reflection of symmetry breaking and remaking

Oleksandr I. Malyi and Alex Zunger

Renewable and Sustainable Energy Institute, University of Colorado, Boulder, Colorado 80309, USA

The YNiO$_3$ nickelate is a paradigm *d*-electron oxide that manifests the intriguing temperature-mediated sequence of three phases transitions from (i) magnetically ordered insulator to (ii) paramagnetic (PM) insulator and then to (iii) PM metal. Such phenomena raised the question of the nature of the association of magnetism and structural symmetry breaking with the appearance in (i) and (ii) and disappearance in (iii) of insulating band gaps. It is demonstrated here that first-principles mean-field-like density functional theory (DFT), driven by molecular dynamics temperature evolution, can describe not only the origin of the magnetically long-range ordered insulating phase (i), but also the creation of an *insulating paramagnet (ii)* that lacks spin- long range order, and of a metallic paramagnet (iii) as temperature rises. This approach provides the patterns of structural and magnetic symmetry breaking at different temperatures, in parallel with band gaps obtained when the evolving geometries are used as input to DFT electronic band structure calculations. This disentangles the complex interplay among spin, charge and orbital degrees of freedom. Analysis shows that the success in describing the rise and fall of the insulating band gaps along the phase transition sequence is enabled by allowing sufficient flexibility in describing diverse local structural and magnetic motifs as input to DFT. This entails the use of sufficiently large supercells that allow expressing structural disproportionation of octahedra, as well as a description of PM phases as a distribution of local magnetic moments (rather than using a single averaged moment). It appears that the historic dismissal of mean-field-like DFT as being unable to describe such Mott-like transitions was premature, as it was based on consideration of averaged crystallographic unit cells, a description that washes out local symmetry-breaking motifs. The magnetically ordered insulating YNiO$_3$ phase (i) and the PM insulating phase (ii) result in DFT from allowing symmetry breaking, evident already by considering the a-thermal internal energy. In contrast, the PM metallic phase (iii) is formed thermally by smearing out thus weakening symmetry breaking. Analysis of snapshots of the different forms of structural vs magnetic symmetry breaking shows that only the loss of the polymorphous distribution of magnetic moments existing in (ii) causes the fall of the band gap, resulting in the metallic state in (iii).The interesting conclusion is that such a description of the rise [ in phases (i) and (ii)] and fall [in phase (iii)] of the insulating gap does not rely on the traditional Mott like strong correlation understanding, but on breaking and remaking of magnetic and structural symmetries reflected in energy lowering.

## I. Introduction

The insulating states both below and above the Néel temperature and the temperature-induced insulator-to-metal transition (IMT) [1] are characteristic features of many 3$d$ Mott oxide systems[2] that have been considered for neuromorphic computing[3] and transparent conductors[4-7]. From a fundamental perspective, the persistent pertinent questions continue to be (a) understanding the nature of the initial insulating phase that emerges from zero gap band degeneracy associated with open-shell configuration [8,9]; (b) understanding the nature of the final metallic phase, which depends on whether the final phase is magnetically ordered or not; (c) understanding the mechanism by which applied "external knobs" (e.g., temperature, pressure, or doping) that transforms a magnetically ordered insulator (i)  or a PM insulator (ii) into a  PM metal (iii) [1,2]. Traditionally, the answers were provided within the context of strongly correlated physics where "electron phases" are the key player. The rise of insulating phases (i) and (ii) from a non-interacting zero gap metal has been conventionally described by electron Coulomb repulsion U (unique to strongly correlated models and absent from mean field-like methods [10-14]) in a largely unresponsive frozen structure. This gap formation mechanism applies both below the Néel temperature to phases with long-range antiferromagnetic (AFM) order and to higher T paramagnetic phases that lack long-range magnetic order. The insulating phases (i) and (ii) were traditionally described as large Hubbard U systems, whereas the metallic phase (iii) has been described [14,15] by a weakening of the U/W Coulomb repulsion U with respect to the bandwidth W. Within such "Electron Phases of Matter theories",  the microscopic degrees of freedom (m-DOF) such as octahedral tilting[16], Jahn–Teller (like) distortion [17], structural/charge disproportionation[18-20], and distribution of non-zero local magnetic moments[8,21-23] are mostly regarded as unresponsive background that is not the cause of the formation of (i)-(iii) but effects  that can be considered later. This viewpoint caused an understandable migration of computational efforts that aim to describe phases (i)-(iii) of open shell $d$-electron oxides (e.g., YNiO$_3$[24-28] and LaTiO$_3$[10,13]) towards strongly correlated methods, leapfrogging mean-field like methods such density functional theory (DFT). However, it remains unclear if DFT was thought to fail in this problem because of its inability to describe strong correlation gap formation in the insulating phases (i) and (ii), or because of the lack of proper description of the local motifs in PM phases (ii) and (iii). For instance, it has been recently demonstrated that accounting for energy-lowering formation of distribution of local spin and structural motifs (i.e., using polymorphous description of quantum material) in 3$d$ Mott oxides can reproduce insulating nature the compounds and experimentally observed symmetry breaking without accounting for any dynamic correction[9,21-23,29]. Interestingly, the distribution of local motifs is also be observed within superposition description[30], Heisenberg Monte-Carlo simulation[31], and Landau–Lifshitz–Gilbert spin dynamics[31] of PM phases of quantum materials.

Herein, we focus on YNiO$_3$ – one of the most representative cases of rare(R)-earth nickelates (RNiO$_3$) that contains all the main features that appear in other systems. Its relevant crystallographic and magnetic configuration are summarized in Table I. They consist of two crystal structures (monoclinic and orthorhombic), two magnetic configurations (antiferromagnetic and paramagnetic), and most notably, the existence of either a single repeated motif ["single local environments" (SLE)] or its disproportionation into two motifs ["Double Local Environments" (DLE)]. The low-temperature monoclinic α phase exhibits both a bond (B) double local environment (B-DLE) and magnetic moment (M) double local environments (M-DLE), meaning it is fully symmetry broken. This monoclinic antiferromagnetic insulator transforms at T~145 K to the β phase being a paramagnetic version of the same monoclinic paramagnetic insulator.[18,32,33] At a high temperature of > 582 K[18], it further transforms to the γ phase being, as far as is known the SLE orthorhombic paramagnetic metal.

**Table I.** Summary of the phases observed of different $YNiO_3$ phases.

| Phase | Crystal space group | Magnetic order | Electronic properties | Temperature (K) | Structural Motif | Magnetic Motif |
|---|---|---|---|---|---|---|
| α | $P2_1/n$ (Monoclinic) | AFM | Insulator | T<145 [18,32,33] | B-DLE | M-DLE |
| β | $P2_1/n$ (Monoclinic) | PM | Insulator | 145<T<582 [18,32,33] | B-DLE | M-DLE |
| γ | Pbnm (Orthorhombic) | PM | Metal | T > 582 K [18] | B-SLE | M-SLE |

We show that the key to the present understanding is that whereas strongly correlated philosophy is needed if the paramagnetic phases (ii) and (iii) are described within small and highly symmetric unit cells that exclude symmetry breaking, relaxing this constrain naturally allows energy-lowering symmetry breaking to come into play. It turns out that DFT (as internal energy minimizer) and DFT molecular dynamics (as free energy equilibration) explain the rise and fall of insulating gaps in $YNiO_3$ paramagnets as a result of structural/magnetic disproportionation and remaking as illustrated in Fig. 1. We also demonstrate that two types of symmetry breakings are most critical: increased temperature causes the loss of structural bond disproportionation (B-DLE becomes B-SLE). However, it is only the loss of magnetic moment distribution (M-DLE becomes M-SLE) that causes the fall of the band gap and rise of the metallic state. With these advances, it is now possible to determine the nature of the insulating state, metallic state, and the transitions between them without depending on pure electronic effects such as strong correlation.

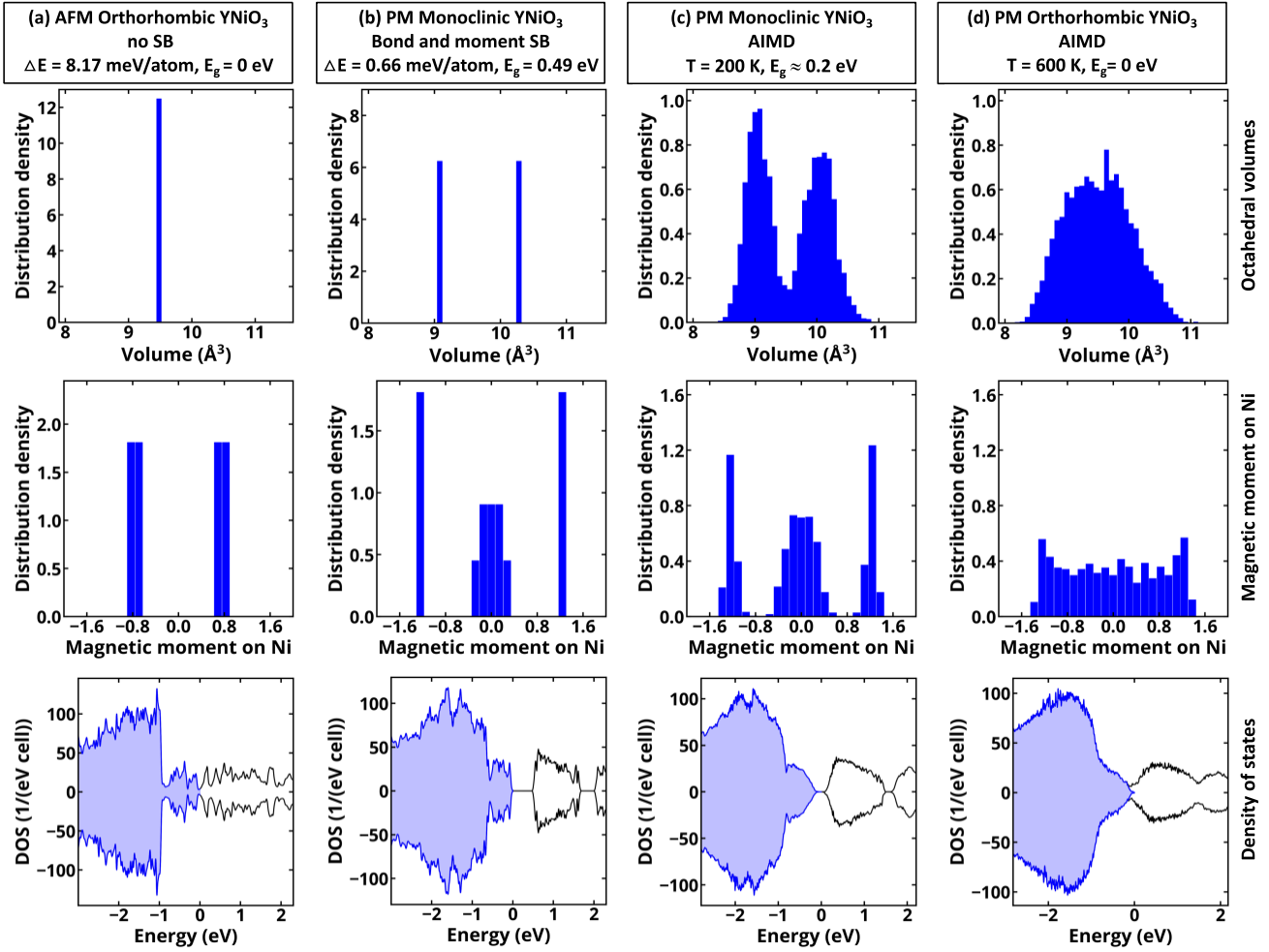

**Figure 1**. Octahedral volume (first raw), local magnetic moments on Ni sublattice (second row), and electronic density of states (third row) from the first-principles calculation of $YNiO_3$. The four columns describe these results as obtained by a sequence of approximations: Column (a) uses a hypothetical monomorphous structure without symmetry breaking (SB), indicating a single characteristic volume and a simple distribution of local moments leading to a metallic zero band gap. Column (b) uses a spin polymorphous structure obtained by replacing an average monomorphous unit cell with a polymorphous supercell structure. This creates a distribution of different octahedral volumes and a non-trivial distribution of local moments, leading to an insulating gap and a significant stabilization in total energy. Column (c) shows *ab initio* molecular dynamics (AIMD) simulation of spin polymorphous lattice at 200 K, and column (d) shows the same as (c) but at 600 K. The symmetry-broken sharp distribution of different unit cell volumes and moments seen in column (b) is broadened at 200 K in Column (c), reducing the band gap. At the temperature of 600 K (Column (d)), thermal displacements overwhelm the tendencies from internal energy minimization, thus completing the Insulator to Metal transition. The averaged density of states (DOS) for occupied states are shown as a shadowed region.

## II. Results

### A. The rise and fall of insulating gaps in $YNiO_3$ as a result of symmetry breaking and remaking (Fig. 1)

(a) When structural symmetry breaking (i.e., formation of B-DLE) is ignored, the orthorhombic AFM (Fig. 1a) phase is a 'false metal' with a single octahedral volume and two opposing magnetic moments.

(b) Allowing in Fig. 1b for full bond and magnetic DLE results in substantial energy lowering (from 8.17 to 0.66 meV/atom) and gives an insulator for the PM monoclinic beta phase. Notably, there is a clear presence of M- and B-DLE – i.e., half of Ni atoms have average magnetic moments of $\pm1.25\mu$ and $0\mu$ with large and small octahedra of distinct volumes. Such symmetry breaking is sufficient to open the gap, as seen in the calculated DOS in Fig. 1b, without the need to involve a dynamic strong correlation.

(c) As Fig. 1c indicates, at 200 K (at which the PM β phase exists experimentally), B-DLE is clearly present in Ab Initio Molecular Dynamics (AIMD), as shown by the district distribution of $NiO_6$ octahedral volumes. These results are similar to those observed based on internal energy minimization for the PM β phase (Fig. 1b) but are broadened, yet preserving the broken symmetry structure. This broadening of local motif distribution is mainly caused by temperature-induced atomic displacements and volume distribution. The broadening of local spin motifs is also reflected in the distribution of magnetic moments, as shown in Fig. 1c. This distribution is similar to that observed by internal energy minimization with some additional temperature broadening.

(d) As Fig. 1d shows, as temperature further increases (e.g., T=600 K, where the PM γ phase exists experimentally), thermal displacements overwhelm the tendencies from internal energy minimization (Fig. 1b), leading to a single Gaussian-like distribution of octahedral volumes and disappearance of the two types of magnetic sublattice (Fig. 1d). As one can see at 200 K (under which PM-β phase exists), the system is an insulator, while at temperature increase to 600 K (i.e., the lowest temperature at which PM γ phase exists experimentally) the system becomes metal.

The results of Fig. 1 explain, from the point of view of simultaneous electronic vs structural and magnetic response, the role of symmetry breaking in the rise and fall of insulating states. Yet, the roles of individual types of symmetry breaking, such as bond disproportionation (B-SLE becomes B-DLE) or spin disproportionation (M-SLE becomes M-DLE) are not explained. In the next section, we will analyze this by calculating the gaps and energy of individual symmetry-breaking snapshots described in section II.B, isolating the critical contributions.

**B. The framework for establishing what hat controls metallization (Fig. 2): the loss of structural bond disproportionation or the loss of magnetic moment disproportionation**

We will find that increased temperature causes the loss of structural bond disproportionation (B-DLE becomes B-SLE). However, it is only the loss of magnetic moment distribution (M-DLE becomes M-SLE) that causes the fall of the band gap and rise of the metallic state. To demonstrate this physics, we create a "string of steps" (1,2,3, etc.) explained in Fig. 2. We start from high-symmetry orthorhombic $YNiO_3$ structure (B-SLE and M-SLE) and allow one additional type of symmetry breaking or temperature at each step. The simplest case shown in Fig. 2(i) is where the octahedra are all identical in shape and volume, and the magnetic moment is also fixed throughout the crystal. This "bond-SLE" as well as "moment-SLE" can thus be represented by repeating as a single repeated motif ("monomorphous"). The next step Fig. 2(ii) illustrates a unit cell made of a mixed B-SLE with M-DLE, whereas Fig. 2(iii) illustrates the case of a bond-DLE (characterized by two different octahedra small and large), and moment-DLE (characterized by two different magnetic motifs—zero or finite moment). Different periodic assemblies of bond and moment SLE and DLE motifs makeup phases of the crystal realized in crystallographic (monoclinic or orthorhombic) and spin (antiferromagnetic and paramagnetic). For example, the fully disproportionated unit cell of B-DLE/M-DLE can exist in an AFM monoclinic phase Fig. 2(iii), or as a PM monoclinic phase Fig. 2(iv), or a PM orthorhombic phase Fig. 2(v), etc.

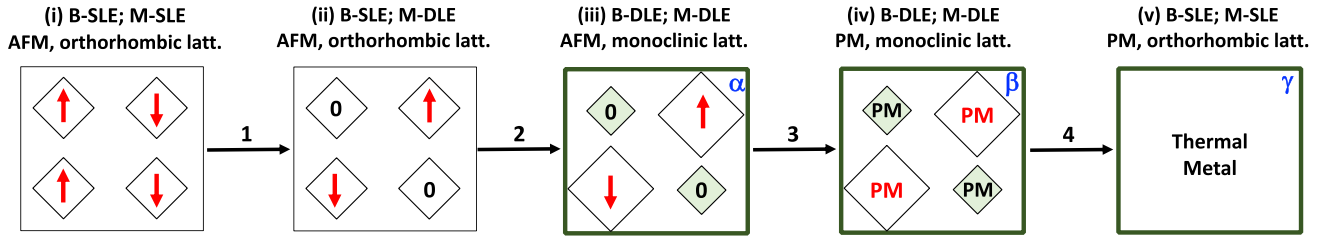

**Figure 2.** Schematic illustration of different moment (M) and bond (B) environments in $YNiO_3$ and transitions between them used to identify the band gap opening and metallization. Bold frames denote physical phases from Table I.

We next describe the first principles results in the sequence of transformation of Fig. 2.

**Transformation 1 (Figs. 2 and 3)**: *The monomorphous assumption consisting of a single structural and magnetic motif [B-SLE/M-SLE], predicts a (false) metal (Fig. 1a). However, magnetic disproportionation leads to gapping even without bond disproportionation (Fig. 3).*

While the observed $\alpha$ phase of $YNiO_3$ has both bond and magnetic disproportionation (i.e., different size $NiO_6$ octahedra and magnetic moments on Ni sublattice) and is an insulator, we wonder if B-DLE is the necessary factor needed to open the gap in $YNiO_3$ relative to the monomorphous state. To answer this question, we have explored the electronic properties of orthorhombic $YNiO_3$ with frozen B-SLE structure, allowing different types of AFM spin configurations. The M-SLE spin configuration has each Ni atom with the same magnetic moments (0.76μ). Since, in this M-SLE, all octahedra have the same spin environment, band gap opening cannot occur. This, however, does not necessarily mean that band gap opening requires bond disproportionation. As shown in Fig. 3, we identify an M-DLE configuration as an insulator with a band gap of 0.26 eV. In such an M-DLE spin configuration, half of the Ni atoms have magnetic moments of 1.06μ while another half has zero magnetic moments, demonstrating that magnetic disproportionation can open the gap even if there is no bond disproportionation.

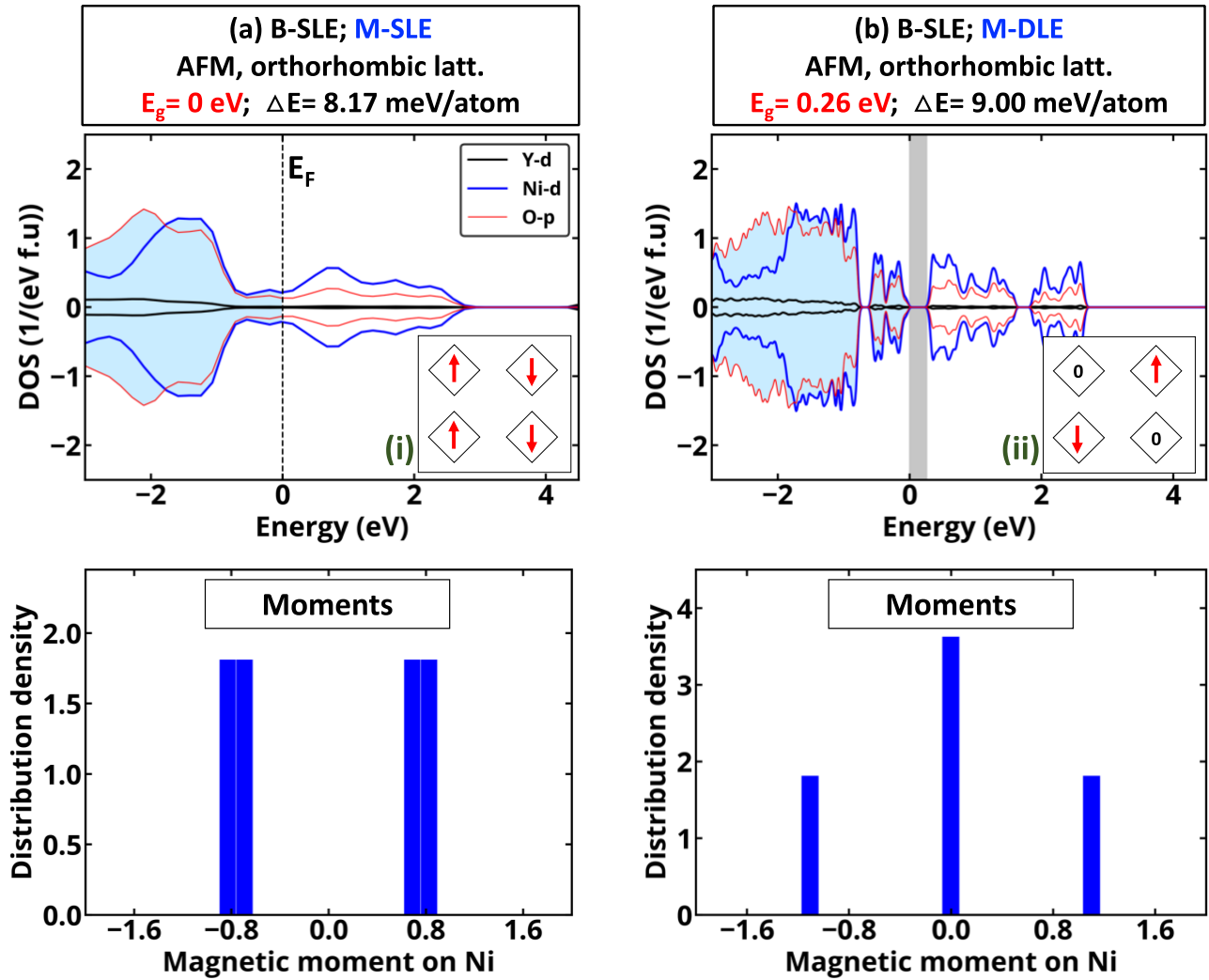

**Figure 3.** Magnetic disproportionation (transition 1) can be the cause of gapping even if there is no bond disproportionation present (i.e., all octahedra are equal). (a) B-SLE/M-SLE monomorphous phase consisting of a single structural and magnetic motif predicts a (false) metal, but (b) magnetic disproportionation alone leads to gapping even without bond disproportionation. Relative energy ($\Delta E$) of different phases is given with respect to AFM $\alpha$ phase.

**Transformation 2 (Figs. 2 and 4):** Bond *disproportionation B-DLE under constant M-DLE enhances the magnitude of the band gap created in the first place by magnetic disproportionation (Fig. 4).*

To understand the role of bond disproportionation on the metal-nonmetal characteristic of YNiO$_3$, we explore the evolution of the band gap as a function of volume difference between large and small octahedra, starting from B-**SL**E/M-DLE transforming into the B-**DLE**/M-DLE $\alpha$ phase. The results are shown in Fig. 4 and demonstrate that octahedral disproportionation is spontaneous, i.e., results in energy lowering. Importantly, the band gap energy is a linear function of the degree of bond disproportionation and is increasing monotonically from 0.26 eV for B-SLE to 0.54 eV for the lowest energy B-DLE structure. Bond disproportionation also affects the magnetic moment on the small octahedra, increasing it from 1.06$\mu$ for B-**SL**E/M-DLE, to 1.23$\mu$ for B-**DLE**/M-DLE. These results thus further confirm that bond disproportion is only an enhancer of the gapping effect caused in the first place by spin disproportionation. We also note that Alonso et al.[18] reported neutron powder diffraction experimental data for the $\alpha$ phase pointing out that the results can be understood as the existence of two Ni sublattices with 1.4 and 0.7μ magnetic moment providing the best fit ($R_{mag}$=15%) for main magnetic reflections. However, the precise measurements of magnetic moments and their special distribution require using the magnetic pair distribution function analysis which was not done in the seminal work of Alonso et al[18] and can further be used to understand spin symmetry breaking in $YNiO_3$ phases.

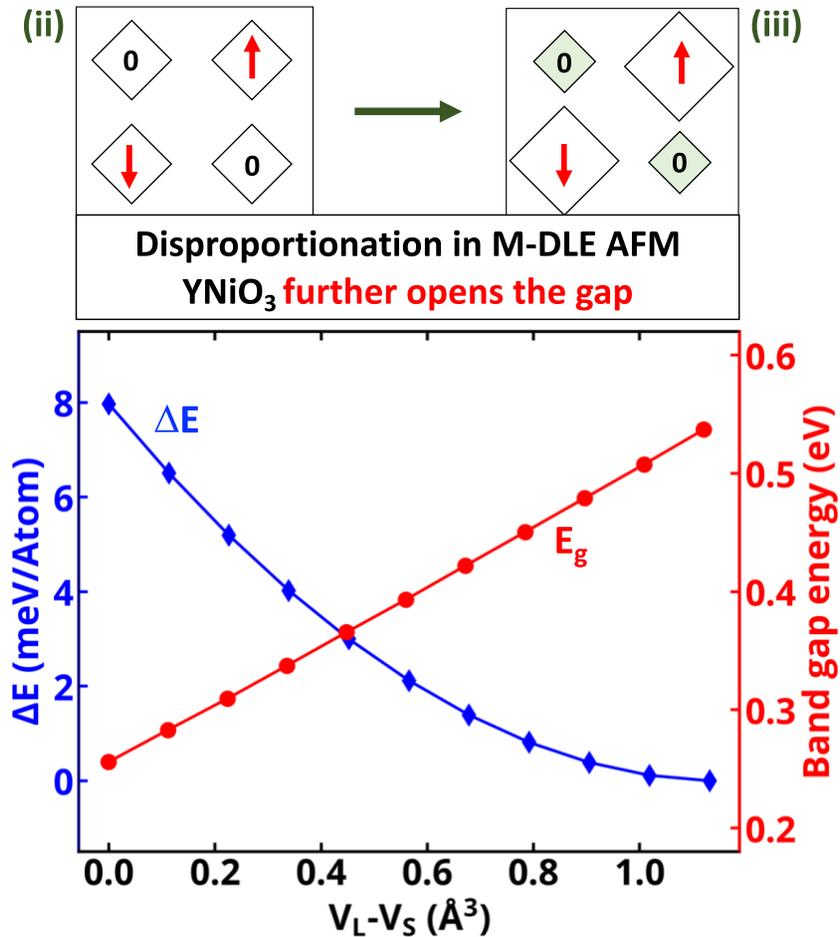

**Figure 4.** Bond disproportionation under constant M-DLE (transition 2) enhances the magnitude of the gap created by magnetic disproportionation. Relative internal energy and band gap energy of AFM M-DLE $YNiO_3$ with orthorhombic lattice vectors as a function of bond disproportionation. The degree of disproportionation is defined as the difference in volumes for large and small $NiO_6$ octahedra (i.e., 0 volume difference corresponds to SLE orthorhombic structure).

**Transformation 3 (Figs. 2 and 5):** *The Neel AFM $\alpha$ to PM $\beta$ transition under constant B-DLE and M-DLE, results in band gap reduction (Fig. 5), but PM $\beta$ is an insulator within mean-field DFT.*

For a long time, it has been believed that mean-field theory cannot describe properties of paramagnetic compounds as such compounds were usually described using global average structure (i.e., *nonmagnetic* model), resulting invariably in a metallic state often conflicting with experimental results. Even in the case of YNiO$_3$, it was argued that DFT cannot do so, indicating, for instance, that "bond-length disproportionation and associated insulating behavior are signatures of a novel correlation effect" [34]. Recently, it has been pointed out [8,21-23,31,35] that band theory can describe paramagnets without appealing for strong correlation by allowing for larger than minimal (super) cells, that permit symmetry breaking and finite local magnetic moments should it lower the total energy. This approach is illustrated here (Fig. 5) for the transition from the α AFM to the β PM phase, indicating the formation of a distribution of local Ni magnetic moments in the PM phase. While the α phase has only two magnetically inequivalent sites, in the β phase, each Ni has its own magnetic moment, which is due to breaking long-range spin order. Such symmetry breaking results in the band gap reduction: the band gap energy for α AFM to β PM phases are 0.59 and 0.49 eV, respectively. These results thus demonstrate that mean-field density functional theory can accurately represent experimentally observed insulator nature of PM β phase and bond disproportionation despite opposite claims in "correlated" literature.

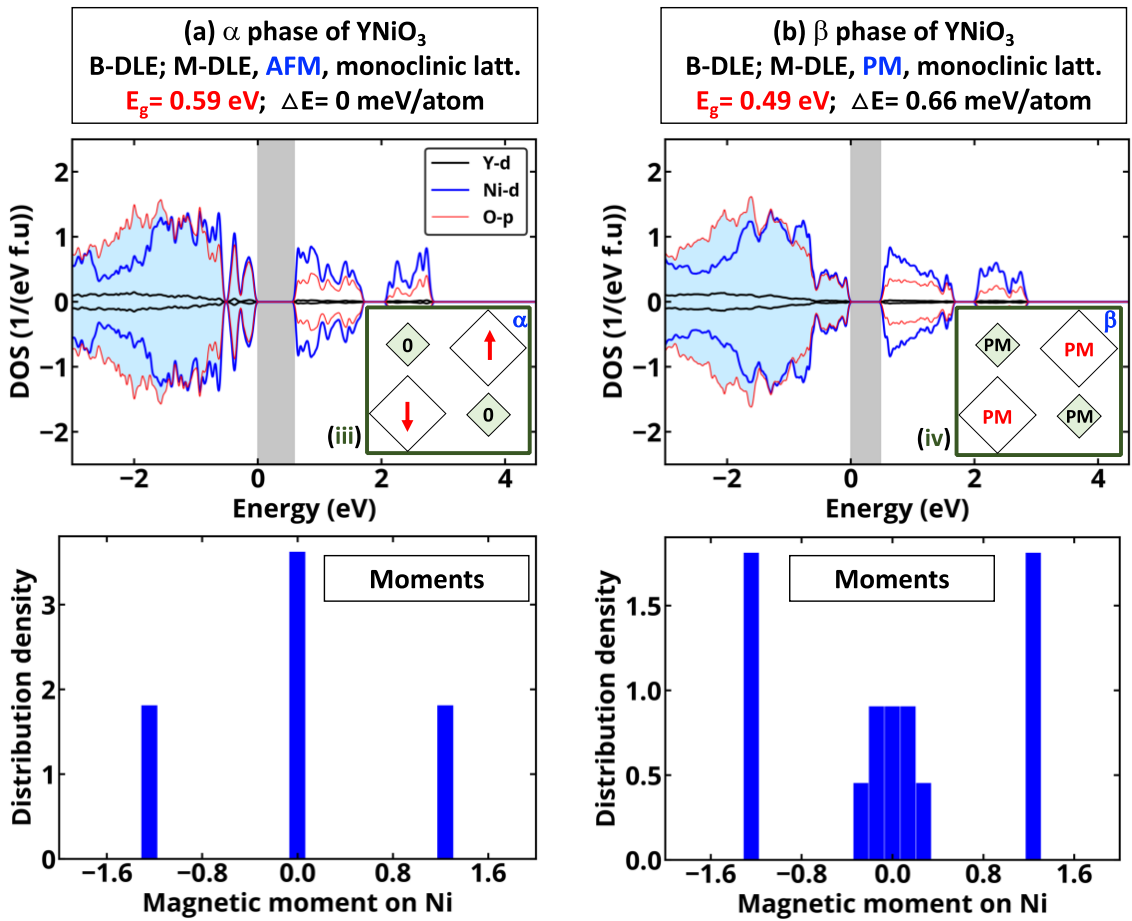

**Figure 5.** The Neel AFM α to PM β monoclinic transition (transition 3) under constant B-DLE and M-DLE results in the band gap reduction. Electronic density of states (top figures) and distribution of local magnetic moments on Ni atoms for (a) α and (b) β phases of YNiO$_3$. For both systems, internal structures and lattice vectors are allowed to relax. The energy difference is given with respect to the alpha phase (i.e., AFM S-DLE M-DLE monoclinic YNiO$_3$).

**Transformation 4 (Fig. 2):** *The PM-β insulating phase transforms to a PM-γ metallic phase at high temperature due to temperature broadening bond and magnetic disproportionation to the point of removing symmetry breaking (Fig. 1d).*

The explanation of transformation 4 of an *insulating* PM-β phase due to the creation of a polymorphous distribution of local moments [8,21,22,35] still faced the need to explain a metallic PM-γ phase at higher temperatures (Table I). One might initially suspect that perhaps the PM-β being insulator, whereas the PM-γ being metallic might have something to do with the former being monoclinic and the latter being orthorhombic. However, orthorhombic PM structure has a spontaneous tendency towards bond and structural disproportionation resulting in band gap opening. Only at high temperatures (Fig. 1d), the internal tendency of the system to energy lowering can be overcome by entropy contribution, which thus eventually results in insulator-metal transition at high temperatures as both bond and magnetic disproportionation are suppressed. We note, however, that even at such high temperatures, the PM-γ phase has a distribution of non-zero magnetic moments, and hence the materials properties cannot be described within the nonmagnetic approximation of the paramagnet.

## III. Discussion

For a long time, band structure models of PM phases of oxides have assumed a single structural motif (e.g., octahedron volume) and a single magnetic motif, leading to the monomorphous model representing an effective "average configuration". Since such model was nonmagnetic, calculations invariably resulted in a metallic state[36,37]. In the many cases where such predicted metallic phases were in conflict with experimental observations, it was argued that these are, in fact, Mott insulators, so the predicted false metal character should be resolved by accounting for strong correlation, codified by the on-site Coulomb repulsion U which splits degenerate bands into occupied and empty. We point out that the existence of a gapped PM phase can be naturally described by mean-field DFT without recourse to a strong correlation. This generally requires that one allows for a larger than a minimal unit cell, enabling structural and/or magnetic symmetry breaking. This leads to a polymorphous structure rather than a virtual averaged structure. Interestingly, $YNiO_3$ has two paramagnetic phases: one at low temperatures being a gapped insulator (β) and one at the higher temperature being ungapped metallic (γ). This poses an interesting corundum as to how those behaviors existing in the same system can be understood. The interesting answer pointed out here is that a gapped insulating phase is evident in first principles mean-field band theory already via minimizing the internal energy alone in an extended supercell. This leads to a polymorphous network having a distribution of both structural and magnetic motifs (whose average is the irrelevant monomorphous approximant). The motif distribution of this a-thermal polymorphous structure is inherited by the beta phase observed in finite temperature molecular dynamics. As the temperature increases, MD shows that the distinct local structural and magnetic motifs lose their polymorphous distribution; when such geometries are used in band theory, a gapless metallic PM gamma phase emerges. Thus, the metallic PM phase emerges from the insulating PM phase not necessarily because of loss of electron correlation but because of thermal motion-induced displacements that alter the electronic band structure.

*Acknowledgment:* The work on magnetic symmetry breaking and electronic structure calculations was supported by U.S. Department of Energy, Office of Science, Basic Energy Sciences, Materials Sciences and Engineering Division within grant DE-SC0010467 to CU Boulder, while using resources of the National Energy Research Scientific Computing Center, which is supported by the Office of Science of the U.S. Department of Energy. The authors also acknowledge the use of computational resources located at the National Renewable Energy Laboratory and sponsored by the Department of Energy's Office of Energy Efficiency and Renewable Energy. Work on structural symmetry breaking and molecular dynamics was supported by NSF-DMREF, grant number 1921949. The authors acknowledge the use of Extreme Science and Engineering Discovery Environment (XSEDE) supercomputer resources, which are supported by the National Science Foundation, grant number ACI-1548562.

## Appendix
**DFT details**

The first-principles calculations are carried out using plane-wave DFT+U as implemented in the Vienna Ab Initio Simulation Package (VASP)[38-40] with PBEsol[41]. For all calculations, a rotationally invariant approach introduced by Dudarev et al. [42] with U-J values of U= 2 eV and J= 0 applied on Ni-d states is utilized. We recall that U in DFT+U acts primarily to reduce the mean field self-interaction in DFT, and thus does not have the same role as U in Hubbard Model (i.e., strong correlation). The cutoff energies for the plane-wave basis are set to 500 eV for final calculations and 550 eV for volume relaxation. Atomic relaxations are performed until the internal forces are smaller than 0.01 eV/Å unless specified. Analysis of structural properties and visualization of computed results are performed using Vesta[43] and pymatgen library[44]. The paramagnetic phase of $YNiO_3$ is simulated by using the *spin* special quasirandom structure (SQS)[45] by decorating a 160-atom supercell with spin up and spin down to create a global zero moment configuration closest to the high-temperature limit of a random spin paramagnet[31]. AIMD simulations are done using the NPT ensemble with an NPT thermostat for 30 ps with the time step of 1 fs. For AIMD simulations, the averaged density of states for N snapshots has been calculated from molecular dynamics snapshots as $DOS = \frac{\Sigma_i^N DOS_i}{N}$, where $DOS_i$ of different snapshots are aligned using O-1s core states. The volume of each octahedron was calculated as the convex hull volume for each octahedron. For each AIMD simulation, 200 snapshots are extracted from molecular dynamics simulations after an equilibration period of 10 ps. The distribution of magnetic moments and density of states are computed by averaging of corresponding quantities for 40 snapshots of ab-initio molecular dynamics simulations after equilibration for 10 ps. The DOS for different snapshots is aligned using different snapshots are aligned using O-1s core states.